\documentclass{emulateapj}
\usepackage{caption2}
\begin{document}
\title{Testing Models for the Shallow Decay Phase of Gamma-Ray Burst Afterglows with Polarization Observations}
\author{Mi-Xiang Lan$^{1,2}$, Xue-Feng Wu$^{3,4}$, and Zi-Gao Dai$^{1,2}$}
\affil{$^{1}$School of Astronomy and Space Science, Nanjing University, Nanjing 210093, China; dzg@nju.edu.cn \\
$^{2}$Key Laboratory of Modern Astronomy and Astrophysics (Nanjing
University), Ministry of Education, China \\
$^{3}$Purple Mountain Observatory, Chinese Academy of Sciences,
Nanjing 210008, China\\
$^{4}$Joint Center for Particle Nuclear Physics and Cosmology of
Purple Mountain Observatory-Nanjing University, Chinese Academy of
Sciences, Nanjing 210008, China}

\begin{abstract}
The X-ray afterglows of almost one half of gamma-ray bursts have been discovered to have a shallow decay phase by the {\em Swift} satellite, whose origin remains mysterious. Two main models have been proposed to explain this phase, relativistic wind bubbles (RWBs) and structured ejecta, which could originate from millisecond magnetars and rapidly-rotating black holes, respectively. Based on these models, we here investigate polarization evolution in the shallow decay phase of X-ray and optical afterglows. We find that in the RWB model, a significant bump of the polarization degree evolution curve appears during the shallow decay phase of both optical and X-ray afterglows, while the polarization position angle changes its direction by $90^\circ$ abruptly. In the structured ejecta model, however, the polarization degree does not evolve significantly during the shallow decay phase of afterglows, no matter whether the magnetic field configuration in the ejecta is random or globally large-scale. Therefore, we conclude that these two models for the shallow decay phase and relevant central engines would be testable with future polarization observations.
\end{abstract}


\section{Introduction}
Thanks to the {\em Swift} satellite (Gehrels et al. 2004), an increasing number of gamma-ray burst (GRB) afterglows have been observed, of which about one half have the shallow decay phase, where the flux density decays as $\propto t^{-\alpha_f}$ with slope of $\alpha_f\sim 0-0.5$. Two popular energy injection models have been proposed so far to explain this phase (for reviews see Zhang 2007 and Kumar \& Zhang 2015). In these models, the injected energy can be either in the form of Poynting flux and/or electron-positron pairs (Dai \& Lu 1998a 1998b; Zhang \& M\'esz\'aros 2001; Dai 2004; Zhang et al. 2006; Yu \& Dai 2007; Dai \& Liu 2012) or in the form of baryons (Rees $\&$ M\'{e}sz\'{a}ros 1998; Sari $\&$ M\'{e}sz\'{a}ros 2000; Nousek et al. 2006), depending on the nature of central engines. On one hand, if the injected energy is initially Poynting flux, a relativistic wind could be dominated by e$^+$e$^-$ pairs at some large radii (Coroniti 1990; Michel 1994; Kirk \& Skj{\ae}raasen 2003; Dai 2004). This e$^+$e$^-$ pair-rich wind collides with the GRB ejecta and a reverse shock occurs. On the other hand, if the injected energy is dominated by baryonic kinetic energy with a wide distribution of bulk Lorentz factor, slower materials eventually catch up with and re-energize the ejecta sweeping up its ambient gas (Rees $\&$ M\'{e}sz\'{a}ros 1998; Sari $\&$ M\'{e}sz\'{a}ros 2000).

Generally speaking, the polarization of emission from a relativistic GRB ejecta depends on the magnetic field configuration, ejecta geometry and structure, and emission mechanism (Shaviv \& Dar 1995; Gruzinov \& Waxman 1999; Eichler \& Levinson 2003; Granot \& K\"{o}nigl 2003; Granot 2003; Lyutikov et al. 2003; Nakar et al. 2003; Dai 2004; Levinson \& Eichler 2004; Lazzati et al. 2004;  Rossi et al. 2004; Wu et al. 2005; Lazzati 2006; Toma et al 2009; Beloborodov 2011; Inoue et al. 2011; Zhang \& Yan 2011; Lan, Wu \& Dai 2016). For the Poynting-flux/$e^\pm$ injection case, i.e. the relativistic wind bubble (RWB) model, a large-scale ordered magnetic field could remain in the wind at large radii, so synchrotron radiation from the shocked wind region would be highly polarized. In this case, the magnetic dipole radiation of a magnetar usually leads to an aligned magnetic field configuration (Spruit et al. 2001).

For the kinetic-energy injection case (i.e. the structured ejecta model), a wind is composed mainly of baryons and leptons (Rees $\&$ M\'{e}sz\'{a}ros 1998; Sari $\&$ M\'{e}sz\'{a}ros 2000). Although whether there is a large-scale ordered magnetic component in the injected energy remains unknown in this case, the X-ray emission detected by the {\em Swift} satellite is usually due to the forward shock emission during energy injection. Because the magnetic field in the forward shock region is commonly assumed to be random, the polarization degree in the X-ray band should be very small. However, if the optical emission is  dominated by the reverse shock emission during energy injection, its polarization evolution could be different from that of the X-ray band.

Several optical polarimeter facilities are now in commission. For example, the Liverpool Telescope (LT; Steele et al. 2004) and the Very Large Telescope (VLT) can detect the polarization evolution of optical afterglows. Thanks to the development of polarimetry detection techniques in X-ray band, a few polarimeter missions are in preparation, e.g., X-ray Timing and Polarimetry (XTP; Jiang et al. 2014), XPOL (Costa et al. 2007), Polarimeters for Energetic Transients (POET; Hill et al. 2008; Bloser et al. 2009), and Gravity and Extreme Magnetism Small explorer (GEMS; Jahoda et al. 2007). Recently, Li et al. (2015) have reported their progress in the X-ray Imaging and Polarimetry Explorer (XIPE), whose systematic error for polarization measurement is less than 1\% at the confidence level of 99\% at 6 keV for the whole sensitive area. Therefore, abundant polarization observations in optical and X-ray bands would be expected in the near future.

In this paper, we investigate polarization evolution during the shallow decay phase of afterglows in both energy bands with two popular energy injection models. We show that these models would be testable with future polarization observations. In our recent paper (Lan, Wu, \& Dai 2016), we calculated polarization evolution of very early optical afterglows and discussed its implications. Our present paper is organized as follows. In Section 2, we discuss polarization evolution during the shallow decay phase of GRB afterglows in both energy bands with the RWB model. In Section 3, we calculate polarization evolution during the shallow decay phase of GRB afterglows with different magnetic field configurations in the structured ejecta model. Finally, in Section 4, we present our conclusions and discussion. As usual, we assume a flat Universe with $\Omega_M=0.27$ and $\Omega_\Lambda=0.73$, and $H_0=71\,{\rm km}\,{\rm s}^{-1}\,{\rm Mpc}^{-1}$. The source is assumed to be located at redshift $z=1$.

\section{Polarization Evolution with the RWB Model}
A rapidly rotating pulsar losses its rotational energy through magnetic dipole radiation, which is in the form of Poynting flux with high magnetization degree $\sigma$ (Michel 1982; Gaensler \& Slane 2006; Hester 2008). However, the observations of pulsar wind nebulae show that a pulsar wind is lepton-dominated (i.e., a low-$\sigma$ outflow; Rees \& Gunn 1974; Kennel \& Coroniti 1984; Begelman \& Li 1992). The physical process of evolution from high $\sigma$ to low $\sigma$ remains unknown (Kargaltsev et al. 2015). A promising process is magnetic reconnection which is induced by  annihilations of reversed magnetic fields near the equatorial plane of an obliquely rotating pulsar between the light cylinder and the termination shock. Such annihilations will lead to an acceleration of leptonic pairs to an ultra-high bulk Lorentz factor at large radii, producing a pulsar wind dominated by electrons and positrons (Michel 1982, 1994; Coroniti 1990; Lyubarsky \& Kirk 2001; Kirk \& Skj{\ae}aasen 2003; Lyubarsky 2003, 2005, 2010a,b; Petri \& Lyubarsky 2007, 2008; Arons 2012; Hoshino \& Lyubarsky 2012). For the Crab nebula, an abrupt acceleration of the pulsar wind is inferred to occur at radii of order $20-50R_{\mathrm{LC}}$ (where $R_{\mathrm{LC}}$ is the light cylinder radius) if the very high energy emission from this nebula is assumed to result from the inverse Compton scattering of pulsed X-ray photons \citep{Aharonian12}. This fact indicates that the pulsar wind becomes lepton-dominated and $\sigma \ll 1$ far within the termination shock.

In the RWB model (Dai 2004), therefore, as a result of the central magnetar activity, Poynting flux is continuously blown out and finally dominated by e$^+$e$^-$ pairs with bulk Lorentz factor of $\gamma_w\sim 10^4-10^7$, as in pulsar wind nebulae (Rees \& Gunn 1974; Kennel \& Coroniti 1984; Begelman \& Li 1992). When this highly relativistic wind catches up with the outer ejecta, two shocks are formed, a forward shock that propagates into the interstellar medium (ISM), and a reverse shock that propagates into the cold wind. So four regions are separated by two shocks: (1) the unshocked ISM, (2) the forward-shocked ISM, (3) the reverse-shocked wind gas, and (4) the unshocked cold wind, where Regions 2 and 3 are separated by a contact discontinuity.

When the injected energy exceeds the initial energy $E_0$ of the outer ejecta, the hydrodynamics of the outer ejecta is changed significantly and the resulting light curve is flattened. Since a large-scale, ordered magnetic field could remain in the wind at large radii, a high polarization degree is predicted in this model during the shallow decay phase. The magnetic field configuration influences the polarization evolution significantly. According to Spruit et al. (2001), a possible magnetic field configuration in the wind injected by a magnetar is aligned. Therefore, we assume that Region 3 has such a configuration and neglect any random magnetic field generated by the reverse shock in this region. In the shocked ISM (Region 2), the magnetic field is assumed to be generated randomly by the forward shock and confined within the shock plane. In the RWB model, therefore, we have the polarization degree $\Pi=[(Q_{\nu,2}+Q_{\nu,3})^2+U_{\nu,3}^2]^{1/2}/(F_{\nu,2}+F_{\nu,3})$ and the position angle $\chi=\frac{1}{2}\arctan(U_{\nu,3}/(Q_{\nu,2}+Q_{\nu,3}))$, where $Q_{\nu,i}$, $U_{\nu,i}$ and $F_{\nu,i}$ are the Stokes parameters of Region i, with $i=2$ for Region 2 and $i=3$ for Region 3 (please note that $U_{\nu,2}=0$, see Lan, Wu \& Dai 2016 for details). The polarization calculation in this paper follows our previous work (Lan, Wu \& Dai 2016). In this paper, we do not assume the polarization degree of power-law electrons in the ordered magnetic field as a constant and we take $\pi_0=\int G(x)N(\gamma_e)d\gamma_e/\int F(x)N(\gamma_e)d\gamma_e$ with $x=\nu'/\nu'_{c}$ (Westfold 1959; Longair 1994; Wu et al. 2005). $F(x)=x\int^\infty_x K_{5/3}(t)dt$, $G(x)=xK_{2/3}(x)$, and $N(\gamma_e)$ is the energy spectrum of electrons. $K_{5/3}(x)$ and $K_{2/3}(x)$ are the modified Bessel functions of $5/3$ and $2/3$ orders. Here $\nu'$ is the observed frequency in the comoving frame of the wind and $\nu'_c$ is the critical frequency of electrons with Lorentz factor $\gamma_e$.

\subsection{Dynamics}
In Dai (2004), the reverse-shocked wind is assumed to be uniform and the structure of the forward-shocked ISM is described by a similarity parameter $\chi$ (Blandford \& McKee 1976). Two critical time scales in the evolution of the system are $T_{M,0}$ and $t_{cr}$. $T_{M,0}$ is the spin-down timescale of the magnetar. At $t_{cr}$ the injected energy is comparable to the initial energy of the ejecta. If $t_{cr} < T_{M,0}$, when $t<t_{cr}$, the energy of the system (including Regions 2, 3 and 4) is dominated by the initial energy $E_0$. With the equality of the velocity and pressure at two sides along the contact discontinuity, the dynamics of the system is described by
\begin{equation}
\gamma_2=\left(\frac{17E_0}{1024\pi n_1m_pc^5t^3}\right)^{1/8},
\end{equation}
and
\begin{equation}
\gamma_3=\left[\frac{(4L_w)^{12/17}(17E_0)^{5/17}}{1024\pi n_1m_pc^5t^{39/17}}\right]^{1/8},
\end{equation}
where $\gamma_2$ and $\gamma_3$ are the bulk Lorentz factors of Regions 2 and 3, $m_p$ is the proton mass, $c$ is the speed of light, $n_1$ is the number density of ISM, $L_w$ is the injected luminosity of the wind, and $t=t_{obs}/(1+z)$, where $t_{obs}$ is the observer's time . With injection of the leptonic wind, the injected energy is comparable to $E_0$ at $t_{cr}$, where the similarity parameter $\chi$ equals to 1. When $t_{cr} < t < T_{M,0}$, the energy increase in Region 2 equals to the work done by Region 3. With the equality of the velocity at two sides along the contact discontinuity, we obtain the dynamical evolution at this stage,
\begin{equation}
\gamma_2=\gamma_3=\left(\frac{L_w}{128\pi n_1m_pc^5t^2}\right)^{1/8}.
\end{equation}
When $t > T_{M,0}$, energy injection from the magnetar becomes unimportant and the evolution of Region 2 is described by the Blandford $\&$ McKee self-similar solution (Blandford $\&$ McKee 1976). So $\gamma_2\propto t^{-3/8}$ and $\gamma_3\propto t^{-7/16}$ (Kobayashi et al. 2000).

\subsection{Polarization evolution}

We numerically calculate the afterglow light curves and polarization evolution in the RWB model. In our calculations, neither the equal arrival time surface effect nor the lateral expansion of the jet is considered. Because the jet opening angle is small (e.g., $\sim 0.1$), the arrival time difference of two photons that are emitted from the center and the edge of the jet is estimated as $\delta t_{\rm obs}\simeq\theta_j^2\times R/2c$. On the other hand, the observer time since the GRB trigger is $t_{\rm obs}\simeq R/2\gamma^2 c$. For a typical shallow decay phase lasting $\sim10^4$ s, we have $\gamma\sim13.7E_{52}^{1/8}n_{1,0}^{-1/8}t_{\rm obs,4}^{-3/8}$ (e.g., Eq.1), which leads to $\delta t_{\rm obs}/t_{\rm obs}\sim(\theta_j\gamma)^2\sim1$. Therefore, this time difference is unimportant as compared to the timescale of the shallow decay phase. In other words, the equal arrival time surface effect will not change the temporal slope of the shallow decay significantly. The lateral expansion of the GRB jet can be neglected in the early times when $\gamma>\theta_j^{-1}$. So we do not need to consider this effect in the shallow decay phase, as the jet phase ($\gamma<\theta_j^{-1}$) is usually later than the shallow decay phase. The distributions of the Lorentz factor and energy density in the wind are assumed to be homogeneous. We take the following parameters: $E_{52}=E_0/10^{52}\,{\rm erg}=0.5$, $n_1=1\,{\rm cm}^{-3}$, $\gamma_w=10^4$, $L_{w,47}=L_w/10^{47}\,{\rm erg}\,{\rm s}^{-1}=36$, the moment of inertia $I_{45}=I/10^{45}\,{\rm g}\,{\rm cm}^2=2.4$, the rotation period $P_0=1\, {\rm ms}$. The energy fractions of electron-positron pairs and ordered magnetic field in the shocked wind are assumed to be $\varepsilon_{e,rs}=0.9$ and $\varepsilon_{B,rs}=0.1$. The spectral index of leptons heated by shocks are $p_{rs}=p_{fs}=2.5$ for Regions 3 and 2. We assume that fractions $\varepsilon_{e,fs}=0.1$ and $\varepsilon_{B,fs}=0.1$ of the internal energy density after the forward shock go to the electrons and magnetic field, respectively. We take the half-opening angle of the wind to be $\theta_j=0.1$.

In Fig. 1, the orientation of the aligned magnetic field is fixed to be $\delta=\pi/4$ and different line styles correspond to different viewing angles ($\theta_V$). When $q\equiv\theta_V/\theta_j=0.0,0.6,1.0$, the shallow decay phase is obvious and during this phase there is a bump of the polarization degree evolution, whose peak corresponds to the ending time of the shallow decay phase. For all of the viewing angles, when the position angle changes abruptly by $90^\circ$, the polarization degree is nonzero. The parameters in Fig. 2 are the same as in Fig. 1, but calculated in optical R-band. The light curves are almost flat until the spin-down time $T_{M,0}$ of the magnetar in optical R-band. The X-ray flux decays before $t_{cr}$, becomes a constant between $t_{cr}$ and $T_{M,0}$, and decays steeper after $T_{M,0}$. We note that the evolution of the polarization degree and position angle in R-band is similar to that in X-ray band.

\begin{figure}
\begin{center}
\includegraphics[width=0.52\textwidth,angle=0]{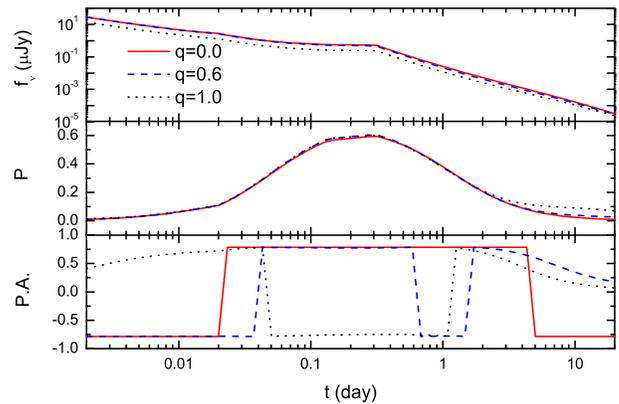}
\caption{Light curves and polarization evolution of $2\,{\rm keV}$ X-ray afterglows in the RWB model. The upper panel shows light curves. The mid panel shows evolution of the polarization degree. The lower panel shows evolution of the position angle. Different line styles correspond to different observing angles for all three panels. The half-opening angle of the wind is $\theta_j=0.1$. The orientation of the ordered magnetic field in Region 3 is $\delta=\pi/4$.} \label{fig1}
\end{center}
\end{figure}

\begin{figure}
\begin{center}
\includegraphics[width=0.52\textwidth,angle=0]{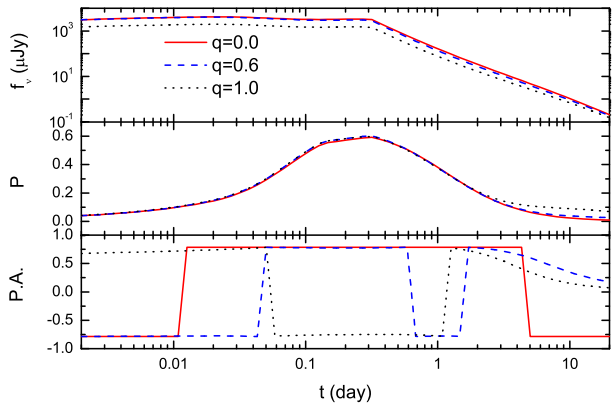}
\caption{Same as Fig. 1 but for optical R-band afterglows.} \label{fig2}
\end{center}
\end{figure}

\section{Polarization Evolution in the Structured Ejecta Model}
An alternative popular model to explain the shallow decay phase in GRB afterglows is the refreshed forward shock, whose energy distribution is a function of the bulk Lorentz factor (Rees $\&$ M\'{e}sz\'{a}ros 1998; Sari $\&$ M\'{e}sz\'{a}ros 2000). In this model, the decelerating ejecta is re-energized by slower ejecta, which leads to a slow decline of the observed flux and hence can explain the shallow decay phase of GRB afterglows.

In this model, the main composition of the slower ejecta is baryons and leptons. However, the magnetic field configuration of the ejecta is uncertain, so we consider two possible cases, i.e. a random field and a toroidal field. And the magnetic field configuration in the forward-shock region is assumed to be random. Here we assume that all the magnetic field is confined in the shock plane. For a random or toroidal magnetic field in the reverse shock region, the polarization degree of emission from the forward-reverse shocked region is $\Pi=(Q_{\nu,2}+Q_{\nu,3})/(F_{\nu,2}+F_{\nu,3})$. We see that when the polarization degree changes from positive to negative or from negative to positive, the position angle changes by $90^\circ$ abruptly.

\subsection{Dynamics}
The ejected mass with Lorentz factor larger than $\gamma$ is assumed to be $M(>\gamma)\propto\gamma^{-s}$ ($s>1$) and the injected energy with this mass is $E(>\gamma)=\gamma Mc^2\equiv E_0(\gamma/\gamma_0)^{-s+1}$, where $E_0$ and $\gamma_0$ are the initial isotropic equivalent energy and initial Lorentz factor respectively. The majority of this injected energy goes into the forward shock region, so we have $E_0(\gamma/\gamma_0)^{-s+1}\simeq E\simeq \gamma^2R^3\rho c^2$. With the assumption that the density of the circum-burst medium is $\rho\propto R^{-g}$, and with the relation $R\simeq 2\gamma^2ct$, we obtain the dynamical evolution of the refreshed shock after the initial deceleration time $t_0$, i.e. $\gamma_2=\gamma_3=\gamma_0(t/t_0)^{-(3-g)/(7+s-2g)}$ and $R=R_0(t/t_0)^{(1+s)/(7+s-2g)}$, where $R_0$ is the deceleration radius of the initial ejecta (where the thin shell approximation is used). When $t<t_0$, we have $\gamma_2=\gamma_0$, $R=2c\gamma_0^2t$ and $E=E_0$. The injection lasts till $t=t_{\rm end}$. When $t>t_{\rm end}$, we find $\gamma_2\propto t^{-3/8}$, $\gamma_3\propto t^{-7/16}$, $R\propto t^{1/4}$ and $E=E(t_{\rm end})$.

\subsection{Polarization evolution}
When $t\leq t_0$, we neglect the reverse shock emission. Because the injected energy before $t_0$ is less than $E_0$ so the injection is unimportant. When $t_0<t\leq t_{\rm end}$, we have $B_{rs}=B_{fs}(\varepsilon_{B,rs}/\varepsilon_{B,fs})^{1/2}$, $\gamma_{m,rs}=\gamma_{m,fs}/\gamma_2(\varepsilon_{e,rs}/\varepsilon_{e,fs})$, $\gamma_{c,rs}=\gamma_{c,fs}(\varepsilon_{B,fs}/\varepsilon_{B,rs})$, and $N_{rs}=N_{fs}\gamma_2$ (Sari $\&$ M\'{e}sz\'{a}ros 2000). Here $B$ is the magnetic field strength, $\gamma_m$ and $\gamma_c$ are the minimum Lorentz factor and the cooling Lorentz factor of the shock-accelerated electrons respectively. The subscripts ``rs'' and ``fs'' denote the quantities for the reverse shocked region and the forward shocked region, respectively. When $t\geq t_{\rm end}$, $\gamma_{m,rs}\propto \gamma_{c,rs}\propto t^{-13/48}$ and $N_{rs}\propto {\rm const}$, and for the forward shock region, we have $\gamma_{m,fs}\propto \gamma_2$ and $\gamma_{c,fs}\propto \gamma_2^{-3}t^{-1}$ for any time $t$ (Sari, Piran $\&$ Narayan 1998).

We then numerically calculate the light curves and polarization evolution in the structured ejecta model with different magnetic field configurations. As in the RWB model, neither the equal arrival time surface effect nor the lateral expansion of the ejecta is considered. The distributions of the Lorentz factor and energy density in the ejecta are assumed to be homogeneous, i.e., these distributions are independent of angle $\theta$ within the ejecta. We take the following parameters: $E_0=10^{52}\,{\rm erg}$, $g=0$, $\gamma_0=2\times10^2$, $n_1=0.1$, $p_{rs}=p_{fs}=2.5$, $\varepsilon_{e,rs}=\varepsilon_{e,fs}=0.02$, $\varepsilon_{B,rs}=0.1$ for an ordered magnetic field, and $\varepsilon_{B,rs}=0.001$ for a random magnetic field, and $\varepsilon_{B,fs}=0.001$. According to the analytic synchrotron spectrum, the appearance of the shallow decay phase requires $s=3p_{fs}-1$ in X-ray band. We also find $\varepsilon_{B,rs}=9/2\sigma$ for the ordered magnetic field, where $\sigma$ is the magnetization degree of the unshocked region. The half-opening angle of the ejecta is 0.1. The injection begins at $t_{0}$ which is the initial deceleration time of the outer ejecta and ends at $t_{\rm end}=0.3\,{\rm day}$.

\begin{figure}
\begin{center}
\includegraphics[width=0.52\textwidth,angle=0]{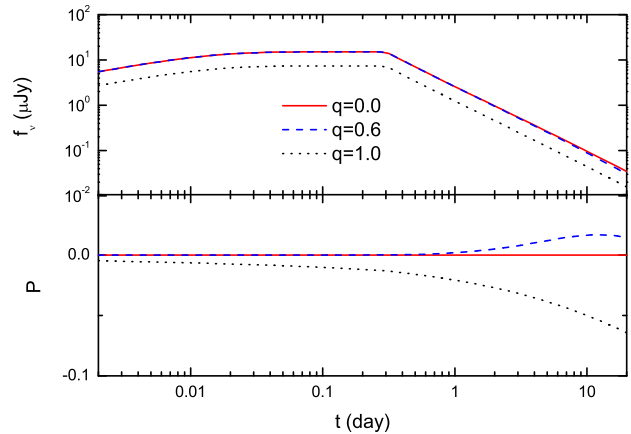}
\caption{Light curves and polarization evolution of $2\,{\rm keV}$ X-ray afterglows in the structured ejecta model, assuming a random magnetic field in the reverse shocked region. The appearance of the shallow decay phase requires $s=3p_{fs}-1$. The upper panel shows the light curves. The lower panel shows evolution of the polarization degree. Different line styles correspond to different observing angles for both panels. The half-opening angle of the ejecta is $\theta_j=0.1$.} \label{fig3}
\end{center}
\end{figure}

\begin{figure}
\begin{center}
\includegraphics[width=0.52\textwidth,angle=0]{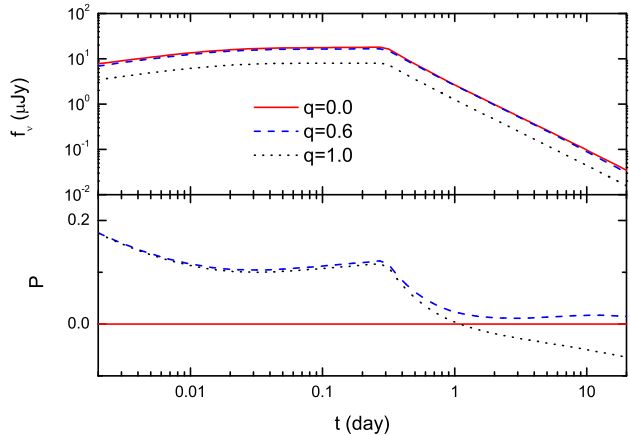}
\caption{Same as Fig. 3 but with a toroidal magnetic field in the reverse shocked region.} \label{fig4}
\end{center}
\end{figure}

Figs. 3 and 4 show light curves and polarization evolution of the structured ejecta in $2\,{\rm keV}$ X-ray band for a random magnetic field and a toroidal magnetic field in the reverse shock region, respectively. Figs. 5 and 6 show the same content as in Figs. 3 and 4, but in optical R-band. The X-ray emission is dominated by the forward shock emission. About $10\%$ polarization degree appears during the X-ray shallow decay phase for the ordered magnetic field configuration (except for $q=0$). The R-band emission is dominated by the reverse shock emission before $t_{\rm end}$ and the polarization degree during the shallow decay phase is about $40\%$ for the ordered magnetic field configuration (except for $q=0$). For a random magnetic field configuration, the polarization degree is approximately zero in both X-ray and R-bands.

\begin{figure}
\begin{center}
\includegraphics[width=0.52\textwidth,angle=0]{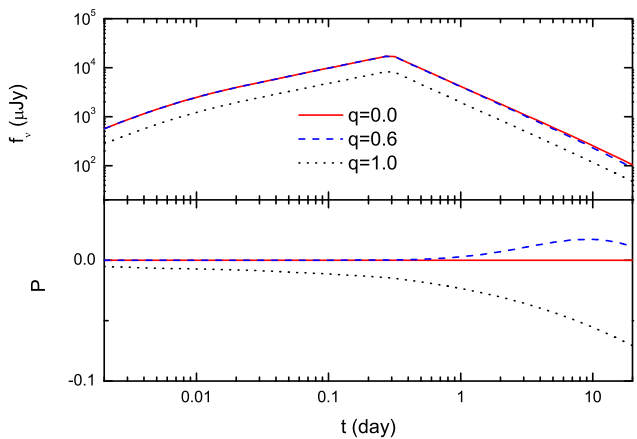}
\caption{Same as Fig. 3 but for optical R-band afterglows.} \label{fig5}
\end{center}
\end{figure}

\begin{figure}
\begin{center}
\includegraphics[width=0.52\textwidth,angle=0]{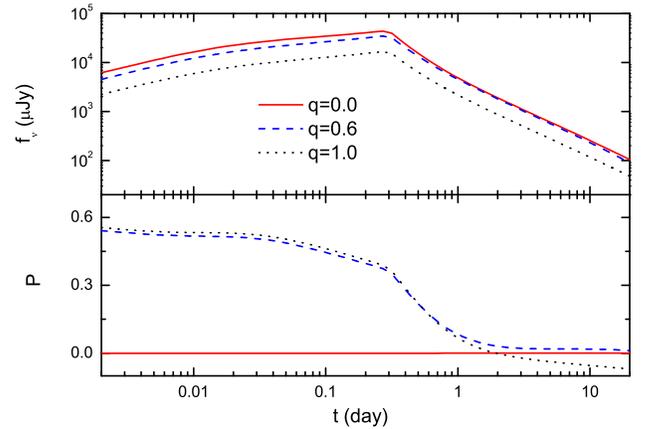}
\caption{Same as Fig. 4 but for optical R-band afterglows.} \label{fig6}
\end{center}
\end{figure}

\section{Conclusions and Discussion}
About one half of GRB X-ray afterglows have the shallow decay phase, which has been widely explained by two energy injection models. Up to now, there have been some optical polarimeters in commission, e.g. LT and VLT. In addition, a few X-ray polarimetry missions are being planed. During the shallow decay phase of GRB afterglows, the emission flux commonly exceeds the detection sensitivities of these detectors and the plateau period is long enough for polarization observations in both X-ray and optical bands. In this paper, we have investigated polarization evolution during the shallow decay phase of GRB afterglows in both the RWB model and the structured ejecta model to show that such two models would be distinguishable with future polarization observations.

In the RWB model, a large-scale ordered magnetic field could be frozen in a relativistic wind. Therefore, synchrotron radiation from the shocked wind region is highly polarized. From our calculations, we found that a bump of the polarization degree evolution indeed appears during the shallow decay phase of optical and X-ray afterglows, whose maximum value can reach about 60\%. In the structured ejecta model, however, the polarization evolution of an afterglow depends on both the magnetic field configuration in the reverse shock region and the ratio of the fluxes of the reverse shock emission to the forward shock emission. For example, a large polarization degree, for an ordered magnetic configuration, would be expected during the optical shallow decay phase only if the reverse shock emission was comparable or dominating over the forward shock emission.



In particular, if an afterglow has the shallow decay phase in either X-ray or optical band, and if non-detection of the polarization degree during this phase occurs, then the structured ejecta model and corresponding black hole engine are preferred. Alternatively, if the shallow decay phase appears in both X-ray and optical bands, and if a bump of the polarization degree evolution is observed and the position angle changes its direction by $90^{\circ}$ abruptly, then the RWB model and relevant magnetar engine are preferred.

\acknowledgements
We thank an anonymous referee for constructive suggestions and Y. F. Huang for useful discussions. This work is supported by the National Basic Research Program (``973'' Program) of China (grant Nos. 2014CB845800 and 2013CB834900) and the National Natural Science Foundation of China (grant Nos. 11573014 and 11322328). X.F.W is also partially supported by the Youth Innovation Promotion Association (2011231), and the Strategic Priority Research Program ``The Emergence of Cosmological Structure'' (grant No. XDB09000000) of the Chinese Academy of Sciences.

\end{document}